\documentclass[aps,twocolumn, floatfix, prl,superscriptaddress]{revtex4-1}
\usepackage{graphicx}
\DeclareGraphicsExtensions{.pdf}
\usepackage{amsmath,amssymb,bbold,bm,color}
\usepackage{float,hyperref}
\usepackage{epstopdf}
\hypersetup{colorlinks=true,urlcolor=blue}

\newcommand{\bk}{{\bm k}}

\newcommand{\bp}{{\bm p}}
\newcommand{\br}{{\bm r}}

\newcommand{\bsig}{{\bm \sigma}}

\newcommand{\cT}{{\cal T}}
\newcommand{\cK}{{\cal K}}

\newcommand{\bee}{\begin{equation}}
\newcommand{\ee}{\end{equation}}

\begin{document}


\title{High-temperature Majorana zero modes}

\author{Alejandro Mercado}
\affiliation{Department of Physics and Astronomy, and Quantum Matter
  Institute, University of British Columbia, Vancouver, BC, Canada V6T 1Z1}
\author{Sharmistha Sahoo}
\affiliation{Department of Physics and Astronomy, and Quantum Matter
  Institute, University of British Columbia, Vancouver, BC, Canada V6T 1Z1}
\affiliation{Department of Physics, Indian Institute of Science, Bengaluru, India 560012}
\author{M. Franz}
\affiliation{Department of Physics and Astronomy, and Quantum Matter
  Institute, University of British Columbia, Vancouver, BC, Canada V6T 1Z1}

\begin{abstract} 
We employ analytical  and numerical approaches to show that
unpaired Majorana zero modes can occur in cores of Abrikosov vortices
at the interface between a three-dimensional topological insulator,
such as Bi$_2$Se$_3$, and a twisted bilayer of high-$T_c$ cuprate
superconductor, such as  Bi$_2$Sr$_2$CaCu$_2$O$_{8+\delta}$. When the
twist angle is close to $45^{\rm o}$ the latter has been predicted to  form a
fully gapped topological superconductor up to temperatures approaching
its native $T_c\simeq 90$K. Majorana zero modes in these structures will persist up to
unprecedented high temperatures and, depending on the quality of the
interface, may be protected by gaps with larger magnitudes than other
candidate systems.  
\end{abstract}

\date{\today}
\maketitle

{\em Introduction --}  Unpaired Majorana zero modes (MZMs) continue
to capture the imagination of
researchers due to their promise in new technologies but also
as an outstanding intellectual challenge in the physics of quantum
matter. An important component in the ongoing worldwide effort
is design and fabrication of new platforms where  MZMs occur robustly under a
wide range of conditions and at accessible temperatures \cite{Alicea2012,Beenakker2012,Leijnse2012,Stanescu2013,Elliott2015,Lutchyn2018}. In this Letter
we discuss a novel platform based on high-$T_c$ cuprate superconductors
where MZMs could persist  up to the native critical temperature of
these materials, which can be as high as 90K in Bi-based
superconductors such as  Bi$_2$Sr$_2$CaCu$_2$O$_{8+\delta}$
(Bi2212). The proposed platform is a variation on the Fu-Kane paradigm \cite{FuKane}
which takes advantage of the anomalous nature of the surface state in
the 3D strong topological insulator (STI) proximitized by an
ordinary superconductor. The Fu-Kane superconductor may have already
been realized in the FeSe family of materials
\cite{Zhang2018,Wang2018,Chen2018,Chiu2020}
although many questions
remain, owing primarily to the complexities associated with the unambiguous MZM
detection which remains a key outstanding challenge \cite{Zhang2018ret}.

An appealing route towards realizing MZMs at higher temperatures is to
employ high-$T_c$ cuprate superconductors. However, the SC order parameter in
all cuprates is known to have a $d_{x^2-y^2}$ symmetry \cite{Tsuei200}
which implies
Dirac nodes in the low-energy quasiparticle excitation spectrum. As a result
MZMs in the proposed setups that involve cuprates
\cite{Li2015,Yan2018,Liu2018,Ortiz2018} would occur on the
background of a sea of gapless Dirac excitations. Unambiguous detection
and technological applications, however, demand MZMs protected by a gap,
preferably with a large amplitude.
In this Letter we propose a strategy to circumvent these inherent
limitations of cuprates by exploiting the recently  discussed
novel physics in {\em twisted} bilayers. According to the
theoretical work \cite{Can2021,volkov2020magic} a 
bilayer assembled from two monolayers of high-$T_c$ cuprate
spontaneously breaks time reversal 
$\cT$ and forms a fully gapped superconductor when the twist angle
$\theta$ is close to 45$^{\rm o}$.  Bi2212 has been demonstrated to
superconduct at high 
temperature in its monolayer form \cite{Yuanbo2019} and is therefore a natural
candidate material for such a bilayer.  The $\cT$-broken state of the bilayer is topological with Chern number $C=2$ or 4
(depending on system parameters) and can be regarded as a
$d_{x^2-y^2}+id_{xy}$ superconductor, or $d+id'$ for short. Lacking significant spin-orbit coupling (SOC), however, the system will not by itself host an MZM in
the core of an Abrikosov vortex \cite{Franz1998}. 

\begin{figure}[t]
\includegraphics[width = 7.0cm]{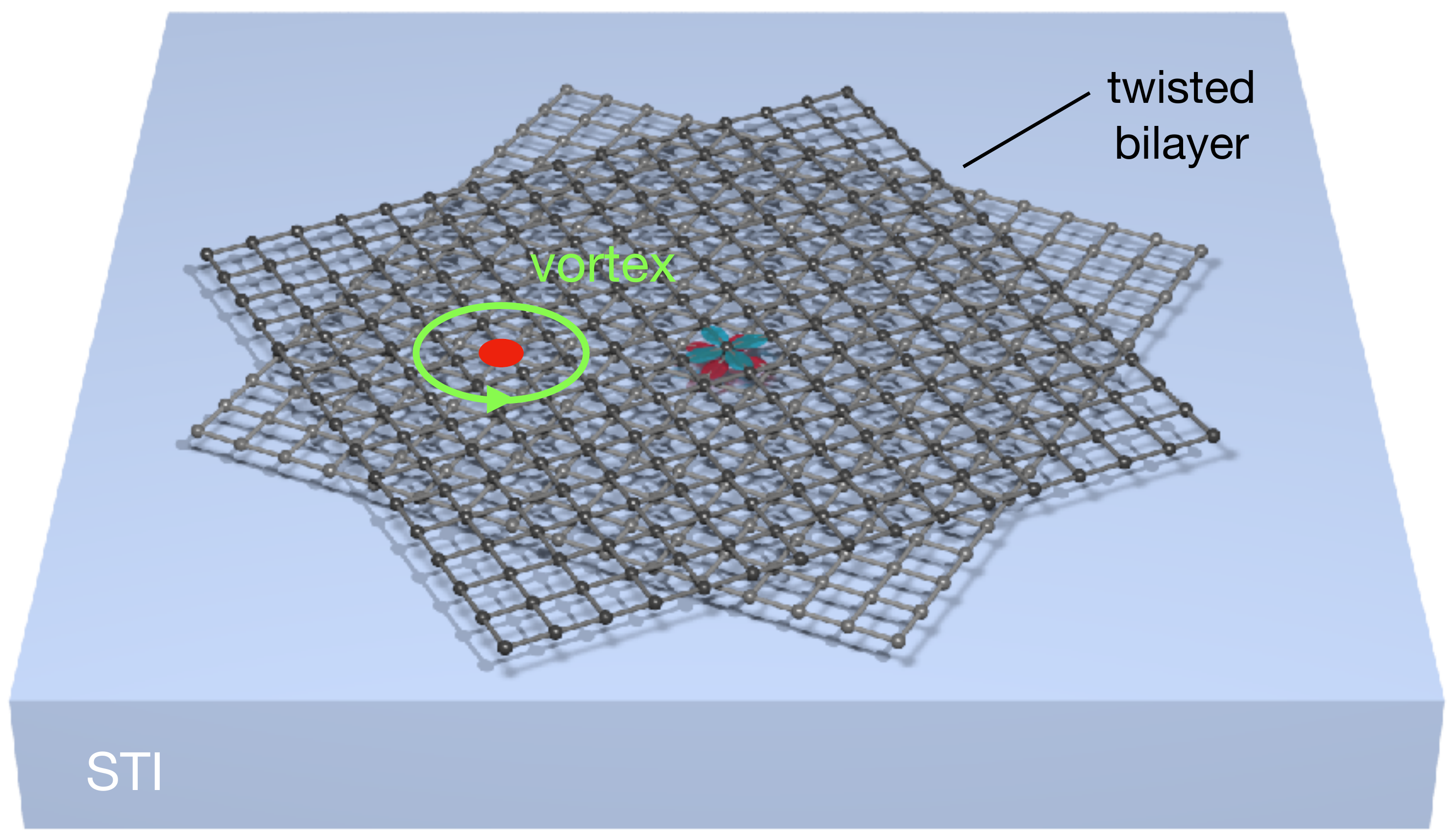}
\caption{Schematic of the proposed setup. A twisted cuprate bilayer, forming a
  $d+id'$ superconducting state, is deposited on the surface of a
  strong topological insulator. 
}\label{fig1}
\end{figure}
Our idea, illustrated in Fig.\ \ref{fig1}, is to use the twisted cuprate
bilayer to proximitize a surface of an STI, which supplies the
requisite SOC.
Owing to the $\cT$-breaking
nature of the $d+id'$ phase and several other factors discussed below,
the mathematical description of the system differs significanly from
the canonical Fu-Kane construction \cite{FuKane} and the existence of
MZMs is not {\em a priori} obvious. 
The two also belong to different symmetry
classes under the Teo-Kane topological classification \cite{Teo2010}.
Nevertheless we demonstrate here  that a simple model describing the
structure in Fig.\ \ref{fig1} hosts an unpaired MZM in the core 
of an Abrikosov vortex. Given the topological stability of isolated
MZMs we argue that their existence is a robust feature of an STI
surface proximitized with a $d+id'$ SC order, insensitive to microscopic details.

Compared to the STI proximitized using a conventional $s$-wave SC our
proposed setup features several key advantages
including potentially larger excitation gaps and
higher critical temperatures. In addition, due to its extremely short
coherence length $\xi$, the cuprate bilayer will exhibit very few
Caroli-de Gennes-Matricon states \cite{Caroli1964} in the vortex core
thus enhancing the MZM  protection from quasiparticle poisoning and
decoherence effects.

{\em The model  --}  The low-energy theory of the STI surface
proximitized with SC order is given by the Bogoliubov-de Gennes (BdG)
Hamiltonian
\begin{equation}\label{h3}
  H_\bk=\begin{pmatrix}
    h_\bk-\mu & \Delta_\bk \\
    \Delta^\dagger_\bk & - h_\bk^\cT+\mu
  \end{pmatrix}.
 \end{equation}
 Here $h_\bk=v(\sigma^yk_x-\sigma^xk_y)$ describes the single massless
 Dirac fermion on the STI surface \cite{HasanKane2010} and 
 $h_\bk^\cT=\sigma^y h_{-\bk}^\ast \sigma^y$ is its time-reversed
counterpart.  $v$ is the surface state velocity,
$\sigma^\alpha$ are Pauli matrices in spin space and $\mu$ denotes the chemical potential. The
$d+id'$ order parameter is defined by 
\begin{equation}\label{h4}
\Delta_\bk=\Delta(k_x^2-k_y^2)+i\Delta'(2k_xk_y)
 \end{equation}
and is, in this representation,
proportional to the unit matrix in spin space. $\Delta$ and $\Delta'$
denote the $d_{x^2-y^2}$ and $d_{xy}$ components, respectively.
The energy eigenvalues can be written as
\begin{equation}\label{h6}
  E_\bk^2=(vk\pm \mu)^2+|\Delta_\bk|^2.
  \end{equation}
 It is important to note that  in view of Eq.\ \eqref{h4} the $d+id'$ order
 parameter yields a fully gapped spectrum only  when $\mu\neq
 0$. As we shall see having to work at non-zero chemical potential is one reason why
solving for the zero mode analytically becomes more difficult than in
the corresponding $s$-wave case where $\Delta_\bk$ is a constant  and one can work at $\mu=0$.

We model a vortex in the SC order parameter by
writing the  Hamiltonian \eqref{h3} in the real-space representation
\begin{equation}\label{h7}
  H=\begin{pmatrix}
    v\bsig\cdot\bp-\mu & \hat\Delta \\
    \hat\Delta^\dagger & -v\bsig\cdot\bp+\mu
  \end{pmatrix},
 \end{equation}
where $\bp=-i\nabla$ is the momentum operator and we performed a uniform
$\pi/2$ rotation around $\sigma^z$ for future convenience.  In order
to enable analytical progress we focus here on the  pure chiral limit
$\Delta=\Delta'$ which, for a single vortex placed at the origin,
yields a  system with full rotation symmetry. The gap
operator can be constructed as described in Ref.\ \cite{vafek2001}
and is given by   
\begin{equation}\label{h8}
  \hat\Delta=a_0^2\left[\{p_+,\{p_+,\Delta(\br)\}\}-
    {i\over 2}\Delta(\br) \left(\partial^2_+\theta(\br)\right)\right].
 \end{equation}
Here $a_0$ is the lattice constant of the underlying lattice model,
 $p_\pm=p_x\pm ip_y$, $\{a,b\}={1\over 2}(ab+ba)$ and
 $\theta(\br)$ denotes the phase of the space dependent gap function
 $\Delta(\br)$. The double symmetrization is required to ensure that
 the resulting operator is hermitian and the last term is needed to
maintain gauge invariance \cite{vafek2001}.

 {\em Vortex core zero mode solution --} The order parameter in the
 presence of a single vortex is given by
\begin{equation}\label{h9}
\Delta(\br)=\Delta_0(r)e^{in\varphi}
 \end{equation}
where $(r,\varphi)$ represent the polar coordinates of vector $\br$  and integer $n$ encodes
vorticity. For the sake of simplicity we assume
$\Delta_0(r)$ to be constant for all $r$, although in
reality the amplitude is required to vanish at the origin.

To find the zero-energy eigenstate $\Psi_0(\br)$ bound to the vortex we follow the strategy
outlined in Ref.\ \cite{cheng2010tunneling}.  The Hamiltonian
\eqref{h7} obeys the antiunitary particle-hole symmetry $\Xi H \Xi=-H$
with $\Xi=\tau^y\sigma^y\cK$, where $\tau^\alpha$ are Pauli matrices in
the Nambu space and $\cK$ denotes complex conjugation. The p-h
symmetry dictates that for every eigenstate of $H$ with energy $E$
there exists an eigenstate $\Psi'=\Xi\Psi$ with energy $-E$. A {\em
  non-degenerate} eigenstate at zero energy must therefore be also an
eigenstate of $\Xi$, that is $\Xi\Psi_0=\lambda\Psi_0$. Furthermore,
because $\Xi^2=1$ it follows that $|\lambda|=1$, and hence
$\lambda=e^{i\beta}$ is a pure phase. When seeking the zero mode we can
  always perform a global U(1) rotation $\Psi_0\to e^{-i\beta/2}
  \Psi_0$ fixing $\lambda=1$, which we will assume to be the case
  henceforth.

If we denote $\Psi_0=(u,v)^T$ with $u$, $v$
  two-component spinors in spin space, it is easy to see that the
  zero-mode condition  $\Xi\Psi_0=\Psi_0$ implies that
  $v=i\sigma^yu^\ast$.  Thus, $\Psi_0$ is fully determined in terms of only two
  independent complex components $u=(u_\uparrow,u_\downarrow)^T$ which
  obey the following equation
\begin{equation}\label{h10}
(v \bsig\cdot\bp-\mu)u+i\sigma^y\hat\Delta u^\ast =0.
 \end{equation}
We solve this equation by passing to polar coordinates and adopting the
ansatz \cite{cheng2010tunneling}
\begin{equation}\label{h11}
  u(r,\varphi)=e^{il\varphi}\begin{pmatrix}
    e^{-i\pi/4}\chi_\uparrow(r) \\
    e^{i(\pi/4+\varphi)}\chi_\downarrow(r)
    \end{pmatrix}
 \end{equation}
with $l$ integer and $\chi_\sigma(r)$ assumed real. Recalling that
$p_\pm=e^{\pm i\varphi}\left(-i\partial_r\pm\partial_\varphi/r\right)$ Eq.\
\eqref{h10} becomes
\begin{eqnarray}\label{h12}
  \mu\chi_\uparrow&-&\left[v\left(\partial_r+{l+1\over r}\right)+\hat\Delta_{l,l+1}
 \right]\chi_\downarrow = 0, \nonumber \\
  \mu\chi_\downarrow&+&\left[v\left(\partial_r-{l\over r}\right)+\hat\Delta_{l+1,l}
  \right]\chi_\uparrow = 0,
\end{eqnarray}
where $\hat\Delta_{l,l'}=e^{-il\varphi}\hat\Delta
e^{-il'\varphi}$. Because $\hat\Delta\sim e^{i(n+2)\varphi}$  the
requirement that  $\hat\Delta_{l,l+1}$ and
$\hat\Delta_{l+1,l}$ are $\varphi$-independent implies $2l+1 =n+2$. It follows
that a non-degenerate  zero-energy solution can exist only when
$l=(n+1)/2$, that is, $l=1$ for a vortex and $l=0$ for
an antivortex.

Working out the required $\hat\Delta_{l,l'}$ operators and passing to a
dimensionless coordinate $\rho=rk_F$ with $k_F=\mu/v$ Eqs.\
\eqref{h12} become, for the $n=1$ vortex,
\begin{eqnarray}\label{h13}
\chi_\uparrow&=&\left[\left(\partial_\rho+{2\over
                 \rho}\right)-\delta\left(\partial^2_\rho+{2\over\rho}\partial_\rho-{1\over
                 4\rho^2}\right)\right]\chi_\downarrow, \nonumber \\
  \chi_\downarrow&=&-\left[\left(\partial_\rho-{1\over \rho}\right)-
                     \delta\left(\partial^2_\rho-{1\over 4\rho^2}\right)
  \right]\chi_\uparrow,
\end{eqnarray}
with $\delta=\Delta_0\mu a_0^2/v^2$ the dimensionless gap
amplitude. It is straightforward to obtain exact numerical solutions of Eqs.\ \eqref{h13}  for an arbitrary gap
parameter $\delta$; an example is given in Fig.\ \ref{fig2}.  We also 
show in Supplementary Material (SM) that a simple analytic form  
\begin{equation}\label{h14}
 \chi_\infty(\rho)=Ae^{-\rho\delta}\begin{pmatrix}
   J_1(\rho) \\
   J_2(\rho)
    \end{pmatrix}
 \end{equation}
solves Eqs.\ \eqref{h13} in the asymptotic region at large $\rho$ for small $\delta$.  As seen in Fig.\ \ref{fig2} 
$\chi_\infty(\rho)$ actually  turns out to be a very good approximation
to the full solution for all distances $\rho$. Similar results can be
established for an antivortex ($n=-1$) and are given in SM along with 
additional details of the calculation.  
\begin{figure}[t]
\includegraphics[width = 7.6cm]{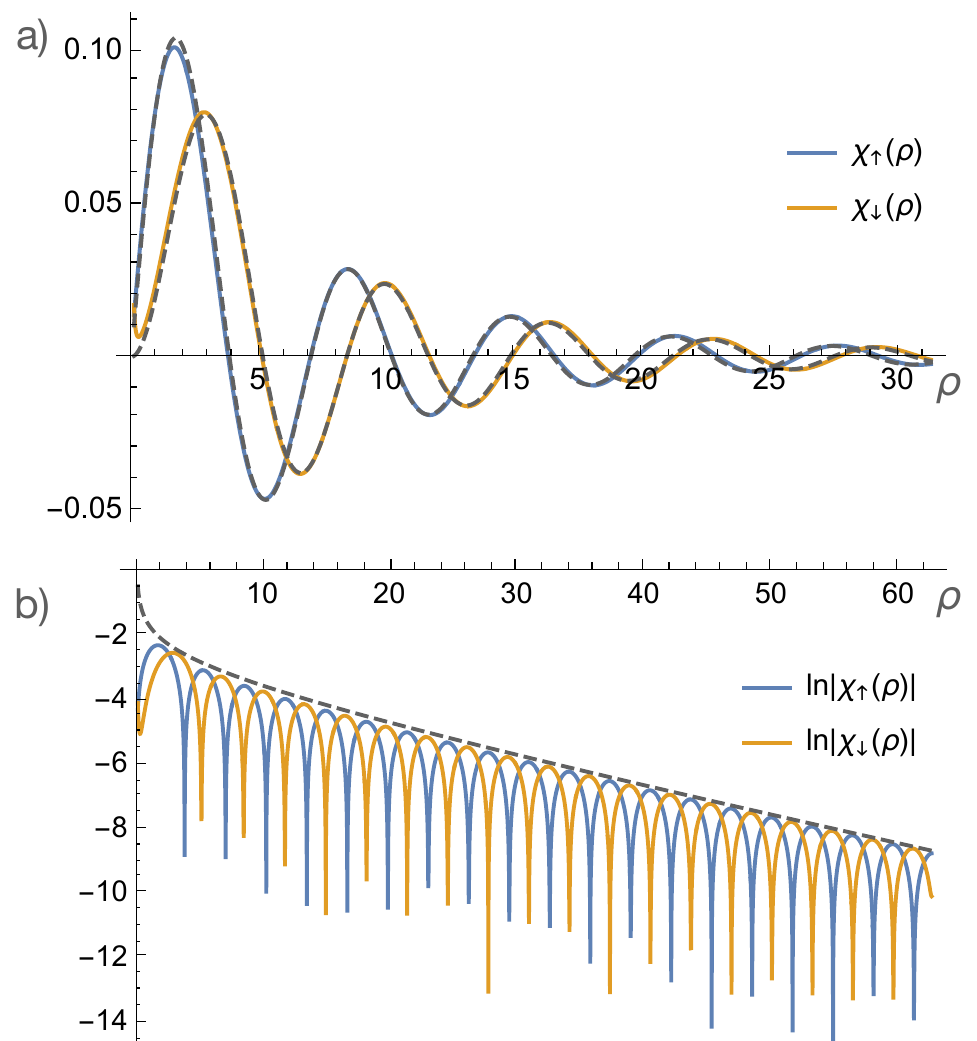}
\caption{Radial components of the zero-mode wavefunction $u(r,\varphi)$ near a vortex. a)
  Normalized solutions obtained by numerical integration of Eqs.\ \eqref{h13}
  with $\delta=0.08$ using the NDSolve algorithm implemented in Wolfram
  Mathematica package (solid lines). Dashed lines represent the
  asymptotic solution $\chi_\infty(\rho)$ given in Eq.\ \eqref{h14}. b) The
  same solutions shown on a logarithmic scale to emphasize the
  exponential decay. The dashed line
  represents the envelope $\sim e^{-\rho\delta}/\sqrt{\rho}$ expected
  on the basis of the long-distance solution Eq.\ \eqref{h14}.
}\label{fig2}
\end{figure}

{\em Lattice model results --}  To confirm the existence of the zero
mode in a broader context we performed numerical simulations within a minimal lattice
model. These do not depend on the assumption of pure chiral limit and
in addition provide useful
information on the entire vortex spectrum, not just the zero mode.  

Although strictly speaking the STI surface cannot be described
by a $\cT$-invariant 2D lattice model \cite{Nielsen1981a,Nielsen1981b}
a workaround exists for situations, such as in the presence
of an Abrikosov vortex, where $\cT$ is explicitly broken.
As demonstrated in Refs.\ \cite{Marchand2012,Pikulin2017} a simple lattice
model can be constructed that faithfully describes the low-energy physics of the
STI surface in the presence of various perturbations, such as
magnetic  or SC coating. The Hamiltonian is given by 
\begin{equation}\label{h18}
h_\bk^{\rm latt}=\lambda(\sigma^y\sin{k_x}-\sigma^x\sin{k_y})+\sigma^zM_\bk,
  \end{equation}
 where $M_\bk=M(2-\cos{k_x}-\cos{k_y})$. The term proportional to
 $\lambda$ is $\cT$-invariant and gives 4 massless Dirac fermions in
 the Brillouin zone. The $M_\bk$ term breaks $\cT$ and leaves only one
 massless Dirac fermion at the $\Gamma$ point $\bk=(0,0)$, gapping out the remaining
 three. Near the $\Gamma$ point the $\cT$-breaking term is small  and
 the model provides a good basis for describing the STI surface state
 in situations where preserving $\cT$ is not essential.
The SC order is captured by defining a lattice
 version of Eq.\ \eqref{h4},
\begin{equation}\label{h19}
 \Delta_\bk^{\rm
   latt}=\Delta(\cos{k_x}-\cos{k_y})+2i\Delta'\sin{k_x}\sin{k_y} 
  \end{equation}
and  the lattice BdG Hamiltonian $H_\bk^{\rm latt}$ is then constructed  following Eq.\
 \eqref{h3}.

\begin{figure}[t]
\includegraphics[width = 8.6cm]{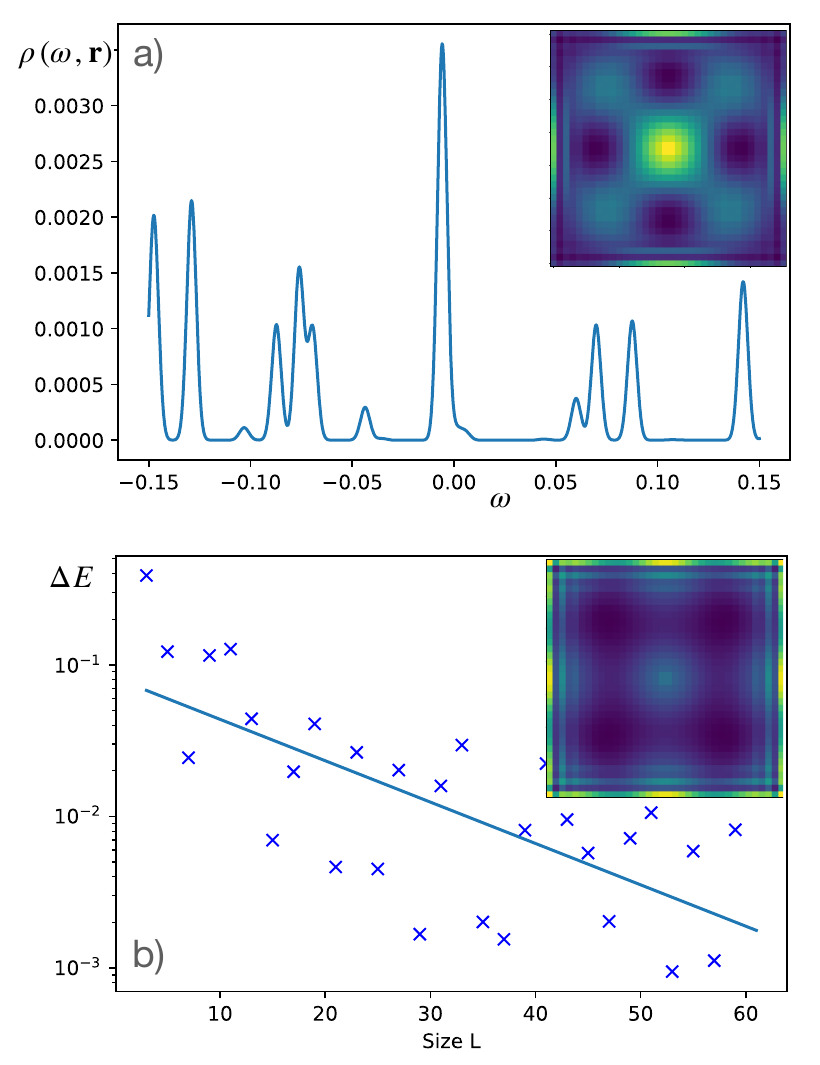}
\caption{Results of the exact diagonalization of the lattice model for a
  single vortex. a) LDOS plotted at the vortex center for $L=35$.  b) Energy
  splitting $\Delta E$ between the lowest energy states whose
  amplitude is shown in the two insets.  Solid line represents a fit
  to $\Delta E(L)=E_0e^{-L/\xi}$ with characteristic lengthscale
  $\xi\simeq 15.8$. Model parameters used, in  
  units of $\lambda$, are
  $(M,\mu,\Delta,\Delta')=(0.5,-0.3,0.15,0.10)$. Many other parameter
  sets were investigated with similar results.
}\label{fig4}
\end{figure}

 To implement the Abrikosov vortex we express
 $H_\bk^{\rm latt}$ in the real-space representation. We consider a
 square sample containing $L\times L$ lattice sites with open boundary
 conditions and  the vortex placed at the center.
 In this representation the BdG Hamiltonian becomes a hermitian
 matrix of size $4L^2$ which we diagonalize numerically for system
 sizes up to $L=59$. More details on the model and its real-space
 implementation in the presence of vortices can be found in Ref.\ \cite{PATHAK2021}.
Fig.\ \ref{fig4} shows the local density of states (LDOS)
\begin{equation}\label{h20}
\rho(\omega,\br)={\sum_n}'\left[|u_n(\br)|^2\delta(\omega-\epsilon_n)+|v_n(\br)|^2\delta(\omega+\epsilon_n)\right],
  \end{equation}
 where the sum is over positive energy eigenvalues $\epsilon_n$ of $H^{\rm latt}$
 and  $\Psi_n(\br)=(u_n(\br),v_n(\br))^T$ are the corresponding 
 eigenstates. When evaluated at the vortex center $\rho(\omega,\br)$
 shows a pronounced peak near zero energy, Fig.\ \ref{fig4}a,
 associated with a pair of eigenvalues $\pm\epsilon_0$ close to zero
 energy, separated by a gap from the rest of the spectrum.
 We identify these as linear combinations of MZMs bound to the
 vortex core and the egde, respectively. Indeed when decomposed into
 $\Xi$ eigenstates using
 the procedure described e.g.\ in Ref.\ \cite{Tarun2021} the
 corresponding  wavefunction amplitudes are localized at the vortex
 core and the edge, respectively (insets Fig.\ \ref{fig4}). The small
 energy splitting is due to the finite size of our sample and should
 decay exponentially with $L$. This is indeed observed in Fig.\ \ref{fig4}b
 where $\Delta E=2\epsilon_0$ is plotted on a semilog scale.

{\em Outlook --} Our results establish the existence of unpaired MZMs
in cores of Abrikosov vortices at the STI surface with an induced
order parameter of $d_{x^2-y^2}+id_{xy}$ symmetry. The latter may
occur in twisted bilayers of high-$T_c$ cuprate superconductors
such as Bi2212 \cite{Can2021}. For the sake of analytical tractability the MZMs were
shown to exist in the very simplest  of models. Nevertheless  their
topological nature ensures stability against {\em arbitrary}
perturbations that do not close the gap.   

While the mechanism leading to the zero mode is
ultimately similar to the classic Fu-Kane construction \cite{FuKane}
it is worth noting that 
the two are {\em not} adiabatically connected and the existence of the MZM in
the $d\pm id'$ case is by no means obvious. From the standpoint of
the general classification of defects \cite{Teo2010} the two
systems belong to different symmetry classes: the former
respects time reversal $\cT$ (in addition to the p-h symmetry
$\Xi$) while in the latter $\cT$ is broken. As discussed in
  SM the  $\cT$ breaking in the present case implies that MZM wavefunctions
  in the vortex and the antivortex are not simply related to one
  another, in stark contrast to the Fu-Kane setup. In addition the fact
  that the   $d+id'$ gap function vanishes at $\bk=0$ implies the necessity
  to work at a non-zero chemical potential $\mu$ which in turn brings
  about the oscillatory nature of the MZM wavefunctions highlighted in
  Fig.\ \ref{fig2}.

Because Bi2212 is known to superconduct up to $90$K in its monolayer form
our proposed construction could open new directions in the pursuit of
Majorana fermions at high temperatures. 
An important measure of the potential usefulness in applications is the
size of the minigap $\Delta_{\rm min}$ protecting the Majorana zero
mode from thermal excitations. In our proposed setup the minigap is
given by the level spacing $\delta E$ between the Caroli-de
Gennes-Matricon vortex core states. In SM we estimate $\delta E\simeq
1/\rho(\mu)\pi R_v^2$ where $\rho(\omega)=\omega/2\pi\hbar^2v^2$ is the
TI surface density of states and   $R_v$ denotes the vortex core
radius, related to the coherence length $\xi$. We show that owing to the small DOS in the TI surface and the short cuprate
coherence length $\xi$ the minigap can be large such that only a handfull of
bound states appear inside the vortex core. Obtaining a
reliable estimate for the minigap size, however, is challenging because it
depends sensitively on the quality of the TI/SC interface as well as the
minimum SC gap in the twisted bilayer. The latter in turn depends on
details of the interlayer coupling at non-zero twist as discussed in
recent works \cite{volkov2021josephson,tummuru2021josephson,Zhao2021,song2021,lu2021}. For these reasons quantitatively accurate estimate of
the minigap and other relevant parameters will likely require
aditional input from experimental studies specifically aimed at
targeting the above uncertainties.

Early work on interfacing topological
insulators with (untwisted) Bi2212 already produced
encouraging results. Mechanically bonded tunnel junctions
\cite{Burch2012} showed evidence of proximity-induced
superconductivity at temperatures up to 80K with an induced SC gap of 13
meV  in Bi$_2$Se$_3$ (and 5 meV in Bi$_2$Te$_3$), a good fraction of the  Bi2212 
maximum gap ($\sim 45$ meV).  Similar
results were obtained by growing Bi$_2$Se$_3$ films on Bi2212 crystals
\cite {Xue2013} where angle-resolved photoemission spectroscopy
revealed an induced
pairing gap $\sim 15$ meV. (We note however, that a subsequent study in a
similar setting failed to detect a measurable gap
\cite{Hasan2014}). These results underscore the urgent need for further
studies of STI/cuprate heterostructures. We also hope  that our
predictions will motivate efforts aimed at fabricating twisted cuprate
bilayers and interfacing them with materials that exhibit strong
spin-orbit coupling,  in search for new platforms
capable of supporting MZMs at elevated temperatures.

{\em Acknowledgments --} The authors are indebted to 
Oguzhan Can, Stephan Plugge and Tarun Tummuru for stimulating discussions. The work
was supported by NSERC and the Canada First Research Excellence Fund,
Quantum Materials and Future Technologies Program.

\bibliography{didsti}

\begin{thebibliography}{40}%
\makeatletter
\providecommand \@ifxundefined [1]{%
 \@ifx{#1\undefined}
}%
\providecommand \@ifnum [1]{%
 \ifnum #1\expandafter \@firstoftwo
 \else \expandafter \@secondoftwo
 \fi
}%
\providecommand \@ifx [1]{%
 \ifx #1\expandafter \@firstoftwo
 \else \expandafter \@secondoftwo
 \fi
}%
\providecommand \natexlab [1]{#1}%
\providecommand \enquote  [1]{``#1''}%
\providecommand \bibnamefont  [1]{#1}%
\providecommand \bibfnamefont [1]{#1}%
\providecommand \citenamefont [1]{#1}%
\providecommand \href@noop [0]{\@secondoftwo}%
\providecommand \href [0]{\begingroup \@sanitize@url \@href}%
\providecommand \@href[1]{\@@startlink{#1}\@@href}%
\providecommand \@@href[1]{\endgroup#1\@@endlink}%
\providecommand \@sanitize@url [0]{\catcode `\\12\catcode `\$12\catcode
  `\&12\catcode `\#12\catcode `\^12\catcode `\_12\catcode `\%12\relax}%
\providecommand \@@startlink[1]{}%
\providecommand \@@endlink[0]{}%
\providecommand \url  [0]{\begingroup\@sanitize@url \@url }%
\providecommand \@url [1]{\endgroup\@href {#1}{\urlprefix }}%
\providecommand \urlprefix  [0]{URL }%
\providecommand \Eprint [0]{\href }%
\providecommand \doibase [0]{http://dx.doi.org/}%
\providecommand \selectlanguage [0]{\@gobble}%
\providecommand \bibinfo  [0]{\@secondoftwo}%
\providecommand \bibfield  [0]{\@secondoftwo}%
\providecommand \translation [1]{[#1]}%
\providecommand \BibitemOpen [0]{}%
\providecommand \bibitemStop [0]{}%
\providecommand \bibitemNoStop [0]{.\EOS\space}%
\providecommand \EOS [0]{\spacefactor3000\relax}%
\providecommand \BibitemShut  [1]{\csname bibitem#1\endcsname}%
\let\auto@bib@innerbib\@empty
\bibitem [{\citenamefont {Alicea}(2012)}]{Alicea2012}%
  \BibitemOpen
  \bibfield  {author} {\bibinfo {author} {\bibfnamefont {J.}~\bibnamefont
  {Alicea}},\ }\href
  {https://iopscience.iop.org/article/10.1088/0034-4885/75/7/076501/meta}
  {\bibfield  {journal} {\bibinfo  {journal} {Rep. Prog. Phys.}\ }\textbf
  {\bibinfo {volume} {75}},\ \bibinfo {pages} {076501} (\bibinfo {year}
  {2012})}\BibitemShut {NoStop}%
\bibitem [{\citenamefont {Beenakker}(2013)}]{Beenakker2012}%
  \BibitemOpen
  \bibfield  {author} {\bibinfo {author} {\bibfnamefont {C.}~\bibnamefont
  {Beenakker}},\ }\href
  {https://www.annualreviews.org/doi/abs/10.1146/annurev-conmatphys-030212-184337}
  {\bibfield  {journal} {\bibinfo  {journal} {Annu. Rev. Con. Mat. Phys.}\
  }\textbf {\bibinfo {volume} {4}},\ \bibinfo {pages} {113} (\bibinfo {year}
  {2013})}\BibitemShut {NoStop}%
\bibitem [{\citenamefont {Leijnse}\ and\ \citenamefont
  {Flensberg}(2012)}]{Leijnse2012}%
  \BibitemOpen
  \bibfield  {author} {\bibinfo {author} {\bibfnamefont {M.}~\bibnamefont
  {Leijnse}}\ and\ \bibinfo {author} {\bibfnamefont {K.}~\bibnamefont
  {Flensberg}},\ }\href
  {https://iopscience.iop.org/article/10.1088/0268-1242/27/12/124003/meta}
  {\bibfield  {journal} {\bibinfo  {journal} {Semiconductor Science and
  Technology}\ }\textbf {\bibinfo {volume} {27}},\ \bibinfo {pages} {124003}
  (\bibinfo {year} {2012})}\BibitemShut {NoStop}%
\bibitem [{\citenamefont {Stanescu}\ and\ \citenamefont
  {Tewari}(2013)}]{Stanescu2013}%
  \BibitemOpen
  \bibfield  {author} {\bibinfo {author} {\bibfnamefont {T.~D.}\ \bibnamefont
  {Stanescu}}\ and\ \bibinfo {author} {\bibfnamefont {S.}~\bibnamefont
  {Tewari}},\ }\href
  {https://iopscience.iop.org/article/10.1088/0953-8984/25/23/233201/meta}
  {\bibfield  {journal} {\bibinfo  {journal} {Journal of Physics: Condensed
  Matter}\ }\textbf {\bibinfo {volume} {25}},\ \bibinfo {pages} {233201}
  (\bibinfo {year} {2013})}\BibitemShut {NoStop}%
\bibitem [{\citenamefont {Elliott}\ and\ \citenamefont
  {Franz}(2015)}]{Elliott2015}%
  \BibitemOpen
  \bibfield  {author} {\bibinfo {author} {\bibfnamefont {S.~R.}\ \bibnamefont
  {Elliott}}\ and\ \bibinfo {author} {\bibfnamefont {M.}~\bibnamefont
  {Franz}},\ }\href
  {https://journals.aps.org/rmp/abstract/10.1103/RevModPhys.87.137} {\bibfield
  {journal} {\bibinfo  {journal} {Rev. Mod. Phys.}\ }\textbf {\bibinfo {volume}
  {87}},\ \bibinfo {pages} {137} (\bibinfo {year} {2015})}\BibitemShut
  {NoStop}%
\bibitem [{\citenamefont {Lutchyn}\ \emph {et~al.}(2018)\citenamefont
  {Lutchyn}, \citenamefont {Bakkers}, \citenamefont {Kouwenhoven},
  \citenamefont {Krogstrup}, \citenamefont {Marcus},\ and\ \citenamefont
  {Oreg}}]{Lutchyn2018}%
  \BibitemOpen
  \bibfield  {author} {\bibinfo {author} {\bibfnamefont {R.~M.}\ \bibnamefont
  {Lutchyn}}, \bibinfo {author} {\bibfnamefont {E.~P. A.~M.}\ \bibnamefont
  {Bakkers}}, \bibinfo {author} {\bibfnamefont {L.~P.}\ \bibnamefont
  {Kouwenhoven}}, \bibinfo {author} {\bibfnamefont {P.}~\bibnamefont
  {Krogstrup}}, \bibinfo {author} {\bibfnamefont {C.~M.}\ \bibnamefont
  {Marcus}}, \ and\ \bibinfo {author} {\bibfnamefont {Y.}~\bibnamefont
  {Oreg}},\ }\href {\doibase 10.1038/s41578-018-0003-1} {\bibfield  {journal}
  {\bibinfo  {journal} {Nature Reviews Materials}\ }\textbf {\bibinfo {volume}
  {3}},\ \bibinfo {pages} {52} (\bibinfo {year} {2018})}\BibitemShut {NoStop}%
\bibitem [{\citenamefont {Fu}\ and\ \citenamefont {Kane}(2008)}]{FuKane}%
  \BibitemOpen
  \bibfield  {author} {\bibinfo {author} {\bibfnamefont {L.}~\bibnamefont
  {Fu}}\ and\ \bibinfo {author} {\bibfnamefont {C.~L.}\ \bibnamefont {Kane}},\
  }\href {\doibase 10.1103/PhysRevLett.100.096407} {\bibfield  {journal}
  {\bibinfo  {journal} {Phys. Rev. Lett.}\ }\textbf {\bibinfo {volume} {100}},\
  \bibinfo {pages} {096407} (\bibinfo {year} {2008})}\BibitemShut {NoStop}%
\bibitem [{\citenamefont {Zhang}\ \emph
  {et~al.}(2018{\natexlab{a}})\citenamefont {Zhang}, \citenamefont {Yaji},
  \citenamefont {Hashimoto}, \citenamefont {Ota}, \citenamefont {Kondo},
  \citenamefont {Okazaki}, \citenamefont {Wang}, \citenamefont {Wen},
  \citenamefont {Gu}, \citenamefont {Ding},\ and\ \citenamefont
  {Shin}}]{Zhang2018}%
  \BibitemOpen
  \bibfield  {author} {\bibinfo {author} {\bibfnamefont {P.}~\bibnamefont
  {Zhang}}, \bibinfo {author} {\bibfnamefont {K.}~\bibnamefont {Yaji}},
  \bibinfo {author} {\bibfnamefont {T.}~\bibnamefont {Hashimoto}}, \bibinfo
  {author} {\bibfnamefont {Y.}~\bibnamefont {Ota}}, \bibinfo {author}
  {\bibfnamefont {T.}~\bibnamefont {Kondo}}, \bibinfo {author} {\bibfnamefont
  {K.}~\bibnamefont {Okazaki}}, \bibinfo {author} {\bibfnamefont
  {Z.}~\bibnamefont {Wang}}, \bibinfo {author} {\bibfnamefont {J.}~\bibnamefont
  {Wen}}, \bibinfo {author} {\bibfnamefont {G.~D.}\ \bibnamefont {Gu}},
  \bibinfo {author} {\bibfnamefont {H.}~\bibnamefont {Ding}}, \ and\ \bibinfo
  {author} {\bibfnamefont {S.}~\bibnamefont {Shin}},\ }\href {\doibase
  10.1126/science.aan4596} {\bibfield  {journal} {\bibinfo  {journal}
  {Science}\ }\textbf {\bibinfo {volume} {360}},\ \bibinfo {pages} {182}
  (\bibinfo {year} {2018}{\natexlab{a}})}\BibitemShut {NoStop}%
\bibitem [{\citenamefont {Wang}\ \emph {et~al.}(2018)\citenamefont {Wang},
  \citenamefont {Kong}, \citenamefont {Fan}, \citenamefont {Chen},
  \citenamefont {Zhu}, \citenamefont {Liu}, \citenamefont {Cao}, \citenamefont
  {Sun}, \citenamefont {Du}, \citenamefont {Schneeloch}, \citenamefont {Zhong},
  \citenamefont {Gu}, \citenamefont {Fu}, \citenamefont {Ding},\ and\
  \citenamefont {Gao}}]{Wang2018}%
  \BibitemOpen
  \bibfield  {author} {\bibinfo {author} {\bibfnamefont {D.}~\bibnamefont
  {Wang}}, \bibinfo {author} {\bibfnamefont {L.}~\bibnamefont {Kong}}, \bibinfo
  {author} {\bibfnamefont {P.}~\bibnamefont {Fan}}, \bibinfo {author}
  {\bibfnamefont {H.}~\bibnamefont {Chen}}, \bibinfo {author} {\bibfnamefont
  {S.}~\bibnamefont {Zhu}}, \bibinfo {author} {\bibfnamefont {W.}~\bibnamefont
  {Liu}}, \bibinfo {author} {\bibfnamefont {L.}~\bibnamefont {Cao}}, \bibinfo
  {author} {\bibfnamefont {Y.}~\bibnamefont {Sun}}, \bibinfo {author}
  {\bibfnamefont {S.}~\bibnamefont {Du}}, \bibinfo {author} {\bibfnamefont
  {J.}~\bibnamefont {Schneeloch}}, \bibinfo {author} {\bibfnamefont
  {R.}~\bibnamefont {Zhong}}, \bibinfo {author} {\bibfnamefont
  {G.}~\bibnamefont {Gu}}, \bibinfo {author} {\bibfnamefont {L.}~\bibnamefont
  {Fu}}, \bibinfo {author} {\bibfnamefont {H.}~\bibnamefont {Ding}}, \ and\
  \bibinfo {author} {\bibfnamefont {H.-J.}\ \bibnamefont {Gao}},\ }\href
  {\doibase 10.1126/science.aao1797} {\bibfield  {journal} {\bibinfo  {journal}
  {Science}\ }\textbf {\bibinfo {volume} {362}},\ \bibinfo {pages} {333}
  (\bibinfo {year} {2018})}\BibitemShut {NoStop}%
\bibitem [{\citenamefont {Chen}\ \emph {et~al.}(2018)\citenamefont {Chen},
  \citenamefont {Chen}, \citenamefont {Yang}, \citenamefont {Du}, \citenamefont
  {Zhu}, \citenamefont {Wang},\ and\ \citenamefont {Wen}}]{Chen2018}%
  \BibitemOpen
  \bibfield  {author} {\bibinfo {author} {\bibfnamefont {M.}~\bibnamefont
  {Chen}}, \bibinfo {author} {\bibfnamefont {X.}~\bibnamefont {Chen}}, \bibinfo
  {author} {\bibfnamefont {H.}~\bibnamefont {Yang}}, \bibinfo {author}
  {\bibfnamefont {Z.}~\bibnamefont {Du}}, \bibinfo {author} {\bibfnamefont
  {X.}~\bibnamefont {Zhu}}, \bibinfo {author} {\bibfnamefont {E.}~\bibnamefont
  {Wang}}, \ and\ \bibinfo {author} {\bibfnamefont {H.-H.}\ \bibnamefont
  {Wen}},\ }\href {\doibase 10.1038/s41467-018-03404-8} {\bibfield  {journal}
  {\bibinfo  {journal} {Nature Communications}\ }\textbf {\bibinfo {volume}
  {9}},\ \bibinfo {pages} {970} (\bibinfo {year} {2018})}\BibitemShut {NoStop}%
\bibitem [{\citenamefont {Chiu}\ \emph {et~al.}(2020)\citenamefont {Chiu},
  \citenamefont {Machida}, \citenamefont {Huang}, \citenamefont {Hanaguri},\
  and\ \citenamefont {Zhang}}]{Chiu2020}%
  \BibitemOpen
  \bibfield  {author} {\bibinfo {author} {\bibfnamefont {C.-K.}\ \bibnamefont
  {Chiu}}, \bibinfo {author} {\bibfnamefont {T.}~\bibnamefont {Machida}},
  \bibinfo {author} {\bibfnamefont {Y.}~\bibnamefont {Huang}}, \bibinfo
  {author} {\bibfnamefont {T.}~\bibnamefont {Hanaguri}}, \ and\ \bibinfo
  {author} {\bibfnamefont {F.-C.}\ \bibnamefont {Zhang}},\ }\href
  {https://advances.sciencemag.org/content/6/9/eaay0443} {\bibfield  {journal}
  {\bibinfo  {journal} {Science Advances}\ }\textbf {\bibinfo {volume} {6}},\
  \bibinfo {pages} {eaay0443} (\bibinfo {year} {2020})}\BibitemShut {NoStop}%
\bibitem [{\citenamefont {Zhang}\ \emph
  {et~al.}(2018{\natexlab{b}})\citenamefont {Zhang}, \citenamefont {Liu},
  \citenamefont {Gazibegovic}, \citenamefont {Xu}, \citenamefont {Logan},
  \citenamefont {Wang}, \citenamefont {van Loo}, \citenamefont {Bommer},
  \citenamefont {de~Moor}, \citenamefont {Car}, \citenamefont {Op~het Veld},
  \citenamefont {van Veldhoven}, \citenamefont {Koelling}, \citenamefont
  {Verheijen}, \citenamefont {Pendharkar}, \citenamefont {Pennachio},
  \citenamefont {Shojaei}, \citenamefont {Lee}, \citenamefont {Palmstr{\o}m},
  \citenamefont {Bakkers}, \citenamefont {Sarma},\ and\ \citenamefont
  {Kouwenhoven}}]{Zhang2018ret}%
  \BibitemOpen
  \bibfield  {author} {\bibinfo {author} {\bibfnamefont {H.}~\bibnamefont
  {Zhang}}, \bibinfo {author} {\bibfnamefont {C.-X.}\ \bibnamefont {Liu}},
  \bibinfo {author} {\bibfnamefont {S.}~\bibnamefont {Gazibegovic}}, \bibinfo
  {author} {\bibfnamefont {D.}~\bibnamefont {Xu}}, \bibinfo {author}
  {\bibfnamefont {J.~A.}\ \bibnamefont {Logan}}, \bibinfo {author}
  {\bibfnamefont {G.}~\bibnamefont {Wang}}, \bibinfo {author} {\bibfnamefont
  {N.}~\bibnamefont {van Loo}}, \bibinfo {author} {\bibfnamefont {J.~D.~S.}\
  \bibnamefont {Bommer}}, \bibinfo {author} {\bibfnamefont {M.~W.~A.}\
  \bibnamefont {de~Moor}}, \bibinfo {author} {\bibfnamefont {D.}~\bibnamefont
  {Car}}, \bibinfo {author} {\bibfnamefont {R.~L.~M.}\ \bibnamefont {Op~het
  Veld}}, \bibinfo {author} {\bibfnamefont {P.~J.}\ \bibnamefont {van
  Veldhoven}}, \bibinfo {author} {\bibfnamefont {S.}~\bibnamefont {Koelling}},
  \bibinfo {author} {\bibfnamefont {M.~A.}\ \bibnamefont {Verheijen}}, \bibinfo
  {author} {\bibfnamefont {M.}~\bibnamefont {Pendharkar}}, \bibinfo {author}
  {\bibfnamefont {D.~J.}\ \bibnamefont {Pennachio}}, \bibinfo {author}
  {\bibfnamefont {B.}~\bibnamefont {Shojaei}}, \bibinfo {author} {\bibfnamefont
  {J.~S.}\ \bibnamefont {Lee}}, \bibinfo {author} {\bibfnamefont {C.~J.}\
  \bibnamefont {Palmstr{\o}m}}, \bibinfo {author} {\bibfnamefont {E.~P. A.~M.}\
  \bibnamefont {Bakkers}}, \bibinfo {author} {\bibfnamefont {S.~D.}\
  \bibnamefont {Sarma}}, \ and\ \bibinfo {author} {\bibfnamefont {L.~P.}\
  \bibnamefont {Kouwenhoven}},\ }\href {\doibase 10.1038/nature26142}
  {\bibfield  {journal} {\bibinfo  {journal} {Nature}\ }\textbf {\bibinfo
  {volume} {556}},\ \bibinfo {pages} {74} (\bibinfo {year}
  {2018}{\natexlab{b}})}\BibitemShut {NoStop}%
\bibitem [{\citenamefont {Tsuei}\ and\ \citenamefont
  {Kirtley}(2000)}]{Tsuei200}%
  \BibitemOpen
  \bibfield  {author} {\bibinfo {author} {\bibfnamefont {C.~C.}\ \bibnamefont
  {Tsuei}}\ and\ \bibinfo {author} {\bibfnamefont {J.~R.}\ \bibnamefont
  {Kirtley}},\ }\href {\doibase 10.1103/RevModPhys.72.969} {\bibfield
  {journal} {\bibinfo  {journal} {Rev. Mod. Phys.}\ }\textbf {\bibinfo {volume}
  {72}},\ \bibinfo {pages} {969} (\bibinfo {year} {2000})}\BibitemShut
  {NoStop}%
\bibitem [{\citenamefont {Li}\ \emph {et~al.}(2015)\citenamefont {Li},
  \citenamefont {Chan},\ and\ \citenamefont {Yao}}]{Li2015}%
  \BibitemOpen
  \bibfield  {author} {\bibinfo {author} {\bibfnamefont {Z.-X.}\ \bibnamefont
  {Li}}, \bibinfo {author} {\bibfnamefont {C.}~\bibnamefont {Chan}}, \ and\
  \bibinfo {author} {\bibfnamefont {H.}~\bibnamefont {Yao}},\ }\href {\doibase
  10.1103/PhysRevB.91.235143} {\bibfield  {journal} {\bibinfo  {journal} {Phys.
  Rev. B}\ }\textbf {\bibinfo {volume} {91}},\ \bibinfo {pages} {235143}
  (\bibinfo {year} {2015})}\BibitemShut {NoStop}%
\bibitem [{\citenamefont {Yan}\ \emph {et~al.}(2018)\citenamefont {Yan},
  \citenamefont {Song},\ and\ \citenamefont {Wang}}]{Yan2018}%
  \BibitemOpen
  \bibfield  {author} {\bibinfo {author} {\bibfnamefont {Z.}~\bibnamefont
  {Yan}}, \bibinfo {author} {\bibfnamefont {F.}~\bibnamefont {Song}}, \ and\
  \bibinfo {author} {\bibfnamefont {Z.}~\bibnamefont {Wang}},\ }\href {\doibase
  10.1103/PhysRevLett.121.096803} {\bibfield  {journal} {\bibinfo  {journal}
  {Phys. Rev. Lett.}\ }\textbf {\bibinfo {volume} {121}},\ \bibinfo {pages}
  {096803} (\bibinfo {year} {2018})}\BibitemShut {NoStop}%
\bibitem [{\citenamefont {Liu}\ \emph {et~al.}(2018)\citenamefont {Liu},
  \citenamefont {He},\ and\ \citenamefont {Nori}}]{Liu2018}%
  \BibitemOpen
  \bibfield  {author} {\bibinfo {author} {\bibfnamefont {T.}~\bibnamefont
  {Liu}}, \bibinfo {author} {\bibfnamefont {J.~J.}\ \bibnamefont {He}}, \ and\
  \bibinfo {author} {\bibfnamefont {F.}~\bibnamefont {Nori}},\ }\href {\doibase
  10.1103/PhysRevB.98.245413} {\bibfield  {journal} {\bibinfo  {journal} {Phys.
  Rev. B}\ }\textbf {\bibinfo {volume} {98}},\ \bibinfo {pages} {245413}
  (\bibinfo {year} {2018})}\BibitemShut {NoStop}%
\bibitem [{\citenamefont {Ortiz}\ \emph {et~al.}(2018)\citenamefont {Ortiz},
  \citenamefont {Varona}, \citenamefont {Viyuela},\ and\ \citenamefont
  {Martin-Delgado}}]{Ortiz2018}%
  \BibitemOpen
  \bibfield  {author} {\bibinfo {author} {\bibfnamefont {L.}~\bibnamefont
  {Ortiz}}, \bibinfo {author} {\bibfnamefont {S.}~\bibnamefont {Varona}},
  \bibinfo {author} {\bibfnamefont {O.}~\bibnamefont {Viyuela}}, \ and\
  \bibinfo {author} {\bibfnamefont {M.~A.}\ \bibnamefont {Martin-Delgado}},\
  }\href {\doibase 10.1103/PhysRevB.97.064501} {\bibfield  {journal} {\bibinfo
  {journal} {Phys. Rev. B}\ }\textbf {\bibinfo {volume} {97}},\ \bibinfo
  {pages} {064501} (\bibinfo {year} {2018})}\BibitemShut {NoStop}%
\bibitem [{\citenamefont {Can}\ \emph {et~al.}(2021)\citenamefont {Can},
  \citenamefont {Tummuru}, \citenamefont {Day}, \citenamefont {Elfimov},
  \citenamefont {Damascelli},\ and\ \citenamefont {Franz}}]{Can2021}%
  \BibitemOpen
  \bibfield  {author} {\bibinfo {author} {\bibfnamefont {O.}~\bibnamefont
  {Can}}, \bibinfo {author} {\bibfnamefont {T.}~\bibnamefont {Tummuru}},
  \bibinfo {author} {\bibfnamefont {R.~P.}\ \bibnamefont {Day}}, \bibinfo
  {author} {\bibfnamefont {I.}~\bibnamefont {Elfimov}}, \bibinfo {author}
  {\bibfnamefont {A.}~\bibnamefont {Damascelli}}, \ and\ \bibinfo {author}
  {\bibfnamefont {M.}~\bibnamefont {Franz}},\ }\href {\doibase
  10.1038/s41567-020-01142-7} {\bibfield  {journal} {\bibinfo  {journal}
  {Nature Physics}\ }\textbf {\bibinfo {volume} {17}},\ \bibinfo {pages} {519}
  (\bibinfo {year} {2021})}\BibitemShut {NoStop}%
\bibitem [{\citenamefont {Volkov}\ \emph {et~al.}(2020)\citenamefont {Volkov},
  \citenamefont {Wilson},\ and\ \citenamefont {Pixley}}]{volkov2020magic}%
  \BibitemOpen
  \bibfield  {author} {\bibinfo {author} {\bibfnamefont {P.~A.}\ \bibnamefont
  {Volkov}}, \bibinfo {author} {\bibfnamefont {J.~H.}\ \bibnamefont {Wilson}},
  \ and\ \bibinfo {author} {\bibfnamefont {J.~H.}\ \bibnamefont {Pixley}},\
  }\href@noop {} {\enquote {\bibinfo {title} {Magic angles and current-induced
  topology in twisted nodal superconductors},}\ } (\bibinfo {year} {2020}),\
  \Eprint {http://arxiv.org/abs/2012.07860} {arXiv:2012.07860
  [cond-mat.supr-con]} \BibitemShut {NoStop}%
\bibitem [{\citenamefont {Yu}\ \emph {et~al.}(2019)\citenamefont {Yu},
  \citenamefont {Ma}, \citenamefont {Cai}, \citenamefont {Zhong}, \citenamefont
  {Ye}, \citenamefont {Shen}, \citenamefont {Gu}, \citenamefont {Chen},\ and\
  \citenamefont {Zhang}}]{Yuanbo2019}%
  \BibitemOpen
  \bibfield  {author} {\bibinfo {author} {\bibfnamefont {Y.}~\bibnamefont
  {Yu}}, \bibinfo {author} {\bibfnamefont {L.}~\bibnamefont {Ma}}, \bibinfo
  {author} {\bibfnamefont {P.}~\bibnamefont {Cai}}, \bibinfo {author}
  {\bibfnamefont {R.}~\bibnamefont {Zhong}}, \bibinfo {author} {\bibfnamefont
  {C.}~\bibnamefont {Ye}}, \bibinfo {author} {\bibfnamefont {J.}~\bibnamefont
  {Shen}}, \bibinfo {author} {\bibfnamefont {G.~D.}\ \bibnamefont {Gu}},
  \bibinfo {author} {\bibfnamefont {X.~H.}\ \bibnamefont {Chen}}, \ and\
  \bibinfo {author} {\bibfnamefont {Y.}~\bibnamefont {Zhang}},\ }\href
  {\doibase 10.1038/s41586-019-1718-x} {\bibfield  {journal} {\bibinfo
  {journal} {Nature}\ }\textbf {\bibinfo {volume} {575}},\ \bibinfo {pages}
  {156} (\bibinfo {year} {2019})}\BibitemShut {NoStop}%
\bibitem [{\citenamefont {Franz}\ and\ \citenamefont {Te\ifmmode \check{s}\else
  \v{s}\fi{}anovi\ifmmode~\acute{c}\else \'{c}\fi{}}(1998)}]{Franz1998}%
  \BibitemOpen
  \bibfield  {author} {\bibinfo {author} {\bibfnamefont {M.}~\bibnamefont
  {Franz}}\ and\ \bibinfo {author} {\bibfnamefont {Z.}~\bibnamefont {Te\ifmmode
  \check{s}\else \v{s}\fi{}anovi\ifmmode~\acute{c}\else \'{c}\fi{}}},\ }\href
  {\doibase 10.1103/PhysRevLett.80.4763} {\bibfield  {journal} {\bibinfo
  {journal} {Phys. Rev. Lett.}\ }\textbf {\bibinfo {volume} {80}},\ \bibinfo
  {pages} {4763} (\bibinfo {year} {1998})}\BibitemShut {NoStop}%
\bibitem [{\citenamefont {Teo}\ and\ \citenamefont {Kane}(2010)}]{Teo2010}%
  \BibitemOpen
  \bibfield  {author} {\bibinfo {author} {\bibfnamefont {J.~C.~Y.}\
  \bibnamefont {Teo}}\ and\ \bibinfo {author} {\bibfnamefont {C.~L.}\
  \bibnamefont {Kane}},\ }\href {\doibase 10.1103/PhysRevB.82.115120}
  {\bibfield  {journal} {\bibinfo  {journal} {Phys. Rev. B}\ }\textbf {\bibinfo
  {volume} {82}},\ \bibinfo {pages} {115120} (\bibinfo {year}
  {2010})}\BibitemShut {NoStop}%
\bibitem [{\citenamefont {Caroli}\ \emph {et~al.}(1964)\citenamefont {Caroli},
  \citenamefont {{De Gennes}},\ and\ \citenamefont {Matricon}}]{Caroli1964}%
  \BibitemOpen
  \bibfield  {author} {\bibinfo {author} {\bibfnamefont {C.}~\bibnamefont
  {Caroli}}, \bibinfo {author} {\bibfnamefont {P.}~\bibnamefont {{De Gennes}}},
  \ and\ \bibinfo {author} {\bibfnamefont {J.}~\bibnamefont {Matricon}},\
  }\href {\doibase https://doi.org/10.1016/0031-9163(64)90375-0} {\bibfield
  {journal} {\bibinfo  {journal} {Physics Letters}\ }\textbf {\bibinfo {volume}
  {9}},\ \bibinfo {pages} {307 } (\bibinfo {year} {1964})}\BibitemShut
  {NoStop}%
\bibitem [{\citenamefont {Hasan}\ and\ \citenamefont
  {Kane}(2010)}]{HasanKane2010}%
  \BibitemOpen
  \bibfield  {author} {\bibinfo {author} {\bibfnamefont {M.~Z.}\ \bibnamefont
  {Hasan}}\ and\ \bibinfo {author} {\bibfnamefont {C.~L.}\ \bibnamefont
  {Kane}},\ }\href {\doibase 10.1103/RevModPhys.82.3045} {\bibfield  {journal}
  {\bibinfo  {journal} {Rev. Mod. Phys.}\ }\textbf {\bibinfo {volume} {82}},\
  \bibinfo {pages} {3045} (\bibinfo {year} {2010})}\BibitemShut {NoStop}%
\bibitem [{\citenamefont {Vafek}\ \emph {et~al.}(2001)\citenamefont {Vafek},
  \citenamefont {Melikyan}, \citenamefont {Franz},\ and\ \citenamefont
  {Te\ifmmode \check{s}\else \v{s}\fi{}anovi\ifmmode~\acute{c}\else
  \'{c}\fi{}}}]{vafek2001}%
  \BibitemOpen
  \bibfield  {author} {\bibinfo {author} {\bibfnamefont {O.}~\bibnamefont
  {Vafek}}, \bibinfo {author} {\bibfnamefont {A.}~\bibnamefont {Melikyan}},
  \bibinfo {author} {\bibfnamefont {M.}~\bibnamefont {Franz}}, \ and\ \bibinfo
  {author} {\bibfnamefont {Z.}~\bibnamefont {Te\ifmmode \check{s}\else
  \v{s}\fi{}anovi\ifmmode~\acute{c}\else \'{c}\fi{}}},\ }\href {\doibase
  10.1103/PhysRevB.63.134509} {\bibfield  {journal} {\bibinfo  {journal} {Phys.
  Rev. B}\ }\textbf {\bibinfo {volume} {63}},\ \bibinfo {pages} {134509}
  (\bibinfo {year} {2001})}\BibitemShut {NoStop}%
\bibitem [{\citenamefont {Cheng}\ \emph {et~al.}(2010)\citenamefont {Cheng},
  \citenamefont {Lutchyn}, \citenamefont {Galitski},\ and\ \citenamefont
  {Sarma}}]{cheng2010tunneling}%
  \BibitemOpen
  \bibfield  {author} {\bibinfo {author} {\bibfnamefont {M.}~\bibnamefont
  {Cheng}}, \bibinfo {author} {\bibfnamefont {R.~M.}\ \bibnamefont {Lutchyn}},
  \bibinfo {author} {\bibfnamefont {V.}~\bibnamefont {Galitski}}, \ and\
  \bibinfo {author} {\bibfnamefont {S.~D.}\ \bibnamefont {Sarma}},\ }\href
  {\doibase 10.1103/physrevb.82.094504} {\bibfield  {journal} {\bibinfo
  {journal} {Physical Review B}\ }\textbf {\bibinfo {volume} {82}},\ \bibinfo
  {pages} {094504} (\bibinfo {year} {2010})}\BibitemShut {NoStop}%
\bibitem [{\citenamefont {Nielsen}\ and\ \citenamefont
  {Ninomiya}(1981)}]{Nielsen1981a}%
  \BibitemOpen
  \bibfield  {author} {\bibinfo {author} {\bibfnamefont {H.~B.}\ \bibnamefont
  {Nielsen}}\ and\ \bibinfo {author} {\bibfnamefont {M.}~\bibnamefont
  {Ninomiya}},\ }\href@noop {} {\bibfield  {journal} {\bibinfo  {journal}
  {Nuclear Physics B}\ }\textbf {\bibinfo {volume} {185}},\ \bibinfo {pages}
  {20} (\bibinfo {year} {1981})}\BibitemShut {NoStop}%
\bibitem [{\citenamefont {Nielsen}\ and\ \citenamefont
  {Ninomya}(1981)}]{Nielsen1981b}%
  \BibitemOpen
  \bibfield  {author} {\bibinfo {author} {\bibfnamefont {H.}~\bibnamefont
  {Nielsen}}\ and\ \bibinfo {author} {\bibfnamefont {M.}~\bibnamefont
  {Ninomya}},\ }\href@noop {} {\bibfield  {journal} {\bibinfo  {journal} {Phys.
  Lett., B}\ }\textbf {\bibinfo {volume} {105}},\ \bibinfo {pages} {219}
  (\bibinfo {year} {1981})}\BibitemShut {NoStop}%
\bibitem [{\citenamefont {Marchand}\ and\ \citenamefont
  {Franz}(2012)}]{Marchand2012}%
  \BibitemOpen
  \bibfield  {author} {\bibinfo {author} {\bibfnamefont {D.~J.~J.}\
  \bibnamefont {Marchand}}\ and\ \bibinfo {author} {\bibfnamefont
  {M.}~\bibnamefont {Franz}},\ }\href {\doibase 10.1103/PhysRevB.86.155146}
  {\bibfield  {journal} {\bibinfo  {journal} {Phys. Rev. B}\ }\textbf {\bibinfo
  {volume} {86}},\ \bibinfo {pages} {155146} (\bibinfo {year}
  {2012})}\BibitemShut {NoStop}%
\bibitem [{\citenamefont {Pikulin}\ and\ \citenamefont
  {Franz}(2017)}]{Pikulin2017}%
  \BibitemOpen
  \bibfield  {author} {\bibinfo {author} {\bibfnamefont {D.~I.}\ \bibnamefont
  {Pikulin}}\ and\ \bibinfo {author} {\bibfnamefont {M.}~\bibnamefont
  {Franz}},\ }\href {\doibase 10.1103/PhysRevX.7.031006} {\bibfield  {journal}
  {\bibinfo  {journal} {Phys. Rev. X}\ }\textbf {\bibinfo {volume} {7}},\
  \bibinfo {pages} {031006} (\bibinfo {year} {2017})}\BibitemShut {NoStop}%
\bibitem [{\citenamefont {Pathak}\ \emph {et~al.}(2021)\citenamefont {Pathak},
  \citenamefont {Plugge},\ and\ \citenamefont {Franz}}]{PATHAK2021}%
  \BibitemOpen
  \bibfield  {author} {\bibinfo {author} {\bibfnamefont {V.}~\bibnamefont
  {Pathak}}, \bibinfo {author} {\bibfnamefont {S.}~\bibnamefont {Plugge}}, \
  and\ \bibinfo {author} {\bibfnamefont {M.}~\bibnamefont {Franz}},\ }\href
  {\doibase https://doi.org/10.1016/j.aop.2021.168431} {\bibfield  {journal}
  {\bibinfo  {journal} {Annals of Physics}\ ,\ \bibinfo {pages} {168431}}
  (\bibinfo {year} {2021})}\BibitemShut {NoStop}%
\bibitem [{\citenamefont {Tummuru}\ \emph {et~al.}(2021)\citenamefont
  {Tummuru}, \citenamefont {Can},\ and\ \citenamefont {Franz}}]{Tarun2021}%
  \BibitemOpen
  \bibfield  {author} {\bibinfo {author} {\bibfnamefont {T.}~\bibnamefont
  {Tummuru}}, \bibinfo {author} {\bibfnamefont {O.}~\bibnamefont {Can}}, \ and\
  \bibinfo {author} {\bibfnamefont {M.}~\bibnamefont {Franz}},\ }\href
  {\doibase 10.1103/PhysRevB.103.L100501} {\bibfield  {journal} {\bibinfo
  {journal} {Phys. Rev. B}\ }\textbf {\bibinfo {volume} {103}},\ \bibinfo
  {pages} {L100501} (\bibinfo {year} {2021})}\BibitemShut {NoStop}%
\bibitem [{\citenamefont {Volkov}\ \emph {et~al.}(2021)\citenamefont {Volkov},
  \citenamefont {Zhao}, \citenamefont {Poccia}, \citenamefont {Cui},
  \citenamefont {Kim},\ and\ \citenamefont {Pixley}}]{volkov2021josephson}%
  \BibitemOpen
  \bibfield  {author} {\bibinfo {author} {\bibfnamefont {P.~A.}\ \bibnamefont
  {Volkov}}, \bibinfo {author} {\bibfnamefont {S.~Y.~F.}\ \bibnamefont {Zhao}},
  \bibinfo {author} {\bibfnamefont {N.}~\bibnamefont {Poccia}}, \bibinfo
  {author} {\bibfnamefont {X.}~\bibnamefont {Cui}}, \bibinfo {author}
  {\bibfnamefont {P.}~\bibnamefont {Kim}}, \ and\ \bibinfo {author}
  {\bibfnamefont {J.}~\bibnamefont {Pixley}},\ }\href@noop {} {\enquote
  {\bibinfo {title} {Josephson effects in twisted nodal superconductors},}\ }
  (\bibinfo {year} {2021}),\ \Eprint {http://arxiv.org/abs/arXiv:2108.13456}
  {arXiv:2108.13456} \BibitemShut {NoStop}%
\bibitem [{\citenamefont {Tummuru}\ \emph {et~al.}(2022)\citenamefont
  {Tummuru}, \citenamefont {Plugge},\ and\ \citenamefont
  {Franz}}]{tummuru2021josephson}%
  \BibitemOpen
  \bibfield  {author} {\bibinfo {author} {\bibfnamefont {T.}~\bibnamefont
  {Tummuru}}, \bibinfo {author} {\bibfnamefont {S.}~\bibnamefont {Plugge}}, \
  and\ \bibinfo {author} {\bibfnamefont {M.}~\bibnamefont {Franz}},\ }\href
  {\doibase 10.1103/PhysRevB.105.064501} {\bibfield  {journal} {\bibinfo
  {journal} {Phys. Rev. B}\ }\textbf {\bibinfo {volume} {105}},\ \bibinfo
  {pages} {064501} (\bibinfo {year} {2022})}\BibitemShut {NoStop}%
\bibitem [{\citenamefont {Zhao}\ \emph {et~al.}(2021)\citenamefont {Zhao},
  \citenamefont {Poccia}, \citenamefont {Cui}, \citenamefont {Volkov},
  \citenamefont {Yoo1}, \citenamefont {Engelke}, \citenamefont {Ronen},
  \citenamefont {Zhong}, \citenamefont {Gu}, \citenamefont {Plugge},
  \citenamefont {Tummuru}, \citenamefont {Franz}, \citenamefont {Pixley},\ and\
  \citenamefont {Kim}}]{Zhao2021}%
  \BibitemOpen
  \bibfield  {author} {\bibinfo {author} {\bibfnamefont {S.~Y.~F.}\
  \bibnamefont {Zhao}}, \bibinfo {author} {\bibfnamefont {N.}~\bibnamefont
  {Poccia}}, \bibinfo {author} {\bibfnamefont {X.}~\bibnamefont {Cui}},
  \bibinfo {author} {\bibfnamefont {P.~A.}\ \bibnamefont {Volkov}}, \bibinfo
  {author} {\bibfnamefont {H.}~\bibnamefont {Yoo1}}, \bibinfo {author}
  {\bibfnamefont {R.}~\bibnamefont {Engelke}}, \bibinfo {author} {\bibfnamefont
  {Y.}~\bibnamefont {Ronen}}, \bibinfo {author} {\bibfnamefont
  {R.}~\bibnamefont {Zhong}}, \bibinfo {author} {\bibfnamefont
  {G.}~\bibnamefont {Gu}}, \bibinfo {author} {\bibfnamefont {S.}~\bibnamefont
  {Plugge}}, \bibinfo {author} {\bibfnamefont {T.}~\bibnamefont {Tummuru}},
  \bibinfo {author} {\bibfnamefont {M.}~\bibnamefont {Franz}}, \bibinfo
  {author} {\bibfnamefont {J.~H.}\ \bibnamefont {Pixley}}, \ and\ \bibinfo
  {author} {\bibfnamefont {P.}~\bibnamefont {Kim}},\ }\href@noop {} {\enquote
  {\bibinfo {title} {Emergent interfacial superconductivity between twisted
  cuprate superconductors},}\ } (\bibinfo {year} {2021}),\ \Eprint
  {http://arxiv.org/abs/2108.13455} {arXiv:2108.13455} \BibitemShut {NoStop}%
\bibitem [{\citenamefont {Song}\ \emph {et~al.}(2021)\citenamefont {Song},
  \citenamefont {Zhang},\ and\ \citenamefont {Vishwanath}}]{song2021}%
  \BibitemOpen
  \bibfield  {author} {\bibinfo {author} {\bibfnamefont {X.-Y.}\ \bibnamefont
  {Song}}, \bibinfo {author} {\bibfnamefont {Y.-H.}\ \bibnamefont {Zhang}}, \
  and\ \bibinfo {author} {\bibfnamefont {A.}~\bibnamefont {Vishwanath}},\
  }\href@noop {} {\enquote {\bibinfo {title} {Doping a moir\'e mott insulator:
  A t-j model study of twisted cuprates},}\ } (\bibinfo {year} {2021}),\
  \Eprint {http://arxiv.org/abs/2109.08142} {arXiv:2109.08142} \BibitemShut
  {NoStop}%
\bibitem [{\citenamefont {Lu}\ and\ \citenamefont {Senechal}(2021)}]{lu2021}%
  \BibitemOpen
  \bibfield  {author} {\bibinfo {author} {\bibfnamefont {X.}~\bibnamefont
  {Lu}}\ and\ \bibinfo {author} {\bibfnamefont {D.}~\bibnamefont {Senechal}},\
  }\href@noop {} {\enquote {\bibinfo {title} {Doping phase diagram of a hubbard
  model for twisted bilayer cuprates},}\ } (\bibinfo {year} {2021}),\ \Eprint
  {http://arxiv.org/abs/2112.00487} {arXiv:2112.00487} \BibitemShut {NoStop}%
\bibitem [{\citenamefont {Zareapour}\ \emph {et~al.}(2012)\citenamefont
  {Zareapour}, \citenamefont {Hayat}, \citenamefont {Zhao}, \citenamefont
  {Kreshchuk}, \citenamefont {Jain}, \citenamefont {Kwok}, \citenamefont {Lee},
  \citenamefont {Cheong}, \citenamefont {Xu}, \citenamefont {Yang},
  \citenamefont {Gu}, \citenamefont {Jia}, \citenamefont {Cava},\ and\
  \citenamefont {Burch}}]{Burch2012}%
  \BibitemOpen
  \bibfield  {author} {\bibinfo {author} {\bibfnamefont {P.}~\bibnamefont
  {Zareapour}}, \bibinfo {author} {\bibfnamefont {A.}~\bibnamefont {Hayat}},
  \bibinfo {author} {\bibfnamefont {S.~Y.~F.}\ \bibnamefont {Zhao}}, \bibinfo
  {author} {\bibfnamefont {M.}~\bibnamefont {Kreshchuk}}, \bibinfo {author}
  {\bibfnamefont {A.}~\bibnamefont {Jain}}, \bibinfo {author} {\bibfnamefont
  {D.~C.}\ \bibnamefont {Kwok}}, \bibinfo {author} {\bibfnamefont
  {N.}~\bibnamefont {Lee}}, \bibinfo {author} {\bibfnamefont {S.-W.}\
  \bibnamefont {Cheong}}, \bibinfo {author} {\bibfnamefont {Z.}~\bibnamefont
  {Xu}}, \bibinfo {author} {\bibfnamefont {A.}~\bibnamefont {Yang}}, \bibinfo
  {author} {\bibfnamefont {G.~D.}\ \bibnamefont {Gu}}, \bibinfo {author}
  {\bibfnamefont {S.}~\bibnamefont {Jia}}, \bibinfo {author} {\bibfnamefont
  {R.~J.}\ \bibnamefont {Cava}}, \ and\ \bibinfo {author} {\bibfnamefont
  {K.~S.}\ \bibnamefont {Burch}},\ }\href {\doibase 10.1038/ncomms2042}
  {\bibfield  {journal} {\bibinfo  {journal} {Nature Communications}\ }\textbf
  {\bibinfo {volume} {3}},\ \bibinfo {pages} {1056} (\bibinfo {year}
  {2012})}\BibitemShut {NoStop}%
\bibitem [{\citenamefont {Wang}\ \emph {et~al.}(2013)\citenamefont {Wang},
  \citenamefont {Ding}, \citenamefont {Fedorov}, \citenamefont {Yao},
  \citenamefont {Li}, \citenamefont {Lv}, \citenamefont {Zhao}, \citenamefont
  {Zhang}, \citenamefont {Xu}, \citenamefont {Schneeloch}, \citenamefont
  {Zhong}, \citenamefont {Ji}, \citenamefont {Wang}, \citenamefont {He},
  \citenamefont {Ma}, \citenamefont {Gu}, \citenamefont {Yao}, \citenamefont
  {Xue}, \citenamefont {Chen},\ and\ \citenamefont {Zhou}}]{Xue2013}%
  \BibitemOpen
  \bibfield  {author} {\bibinfo {author} {\bibfnamefont {E.}~\bibnamefont
  {Wang}}, \bibinfo {author} {\bibfnamefont {H.}~\bibnamefont {Ding}}, \bibinfo
  {author} {\bibfnamefont {A.~V.}\ \bibnamefont {Fedorov}}, \bibinfo {author}
  {\bibfnamefont {W.}~\bibnamefont {Yao}}, \bibinfo {author} {\bibfnamefont
  {Z.}~\bibnamefont {Li}}, \bibinfo {author} {\bibfnamefont {Y.-F.}\
  \bibnamefont {Lv}}, \bibinfo {author} {\bibfnamefont {K.}~\bibnamefont
  {Zhao}}, \bibinfo {author} {\bibfnamefont {L.-G.}\ \bibnamefont {Zhang}},
  \bibinfo {author} {\bibfnamefont {Z.}~\bibnamefont {Xu}}, \bibinfo {author}
  {\bibfnamefont {J.}~\bibnamefont {Schneeloch}}, \bibinfo {author}
  {\bibfnamefont {R.}~\bibnamefont {Zhong}}, \bibinfo {author} {\bibfnamefont
  {S.-H.}\ \bibnamefont {Ji}}, \bibinfo {author} {\bibfnamefont
  {L.}~\bibnamefont {Wang}}, \bibinfo {author} {\bibfnamefont {K.}~\bibnamefont
  {He}}, \bibinfo {author} {\bibfnamefont {X.}~\bibnamefont {Ma}}, \bibinfo
  {author} {\bibfnamefont {G.}~\bibnamefont {Gu}}, \bibinfo {author}
  {\bibfnamefont {H.}~\bibnamefont {Yao}}, \bibinfo {author} {\bibfnamefont
  {Q.-K.}\ \bibnamefont {Xue}}, \bibinfo {author} {\bibfnamefont
  {X.}~\bibnamefont {Chen}}, \ and\ \bibinfo {author} {\bibfnamefont
  {S.}~\bibnamefont {Zhou}},\ }\href {\doibase 10.1038/nphys2744} {\bibfield
  {journal} {\bibinfo  {journal} {Nature Physics}\ }\textbf {\bibinfo {volume}
  {9}},\ \bibinfo {pages} {621} (\bibinfo {year} {2013})}\BibitemShut {NoStop}%
\bibitem [{\citenamefont {Xu}\ \emph {et~al.}(2014)\citenamefont {Xu},
  \citenamefont {Liu}, \citenamefont {Richardella}, \citenamefont {Belopolski},
  \citenamefont {Alidoust}, \citenamefont {Neupane}, \citenamefont {Bian},
  \citenamefont {Samarth},\ and\ \citenamefont {Hasan}}]{Hasan2014}%
  \BibitemOpen
  \bibfield  {author} {\bibinfo {author} {\bibfnamefont {S.-Y.}\ \bibnamefont
  {Xu}}, \bibinfo {author} {\bibfnamefont {C.}~\bibnamefont {Liu}}, \bibinfo
  {author} {\bibfnamefont {A.}~\bibnamefont {Richardella}}, \bibinfo {author}
  {\bibfnamefont {I.}~\bibnamefont {Belopolski}}, \bibinfo {author}
  {\bibfnamefont {N.}~\bibnamefont {Alidoust}}, \bibinfo {author}
  {\bibfnamefont {M.}~\bibnamefont {Neupane}}, \bibinfo {author} {\bibfnamefont
  {G.}~\bibnamefont {Bian}}, \bibinfo {author} {\bibfnamefont {N.}~\bibnamefont
  {Samarth}}, \ and\ \bibinfo {author} {\bibfnamefont {M.~Z.}\ \bibnamefont
  {Hasan}},\ }\href {\doibase 10.1103/PhysRevB.90.085128} {\bibfield  {journal}
  {\bibinfo  {journal} {Phys. Rev. B}\ }\textbf {\bibinfo {volume} {90}},\
  \bibinfo {pages} {085128} (\bibinfo {year} {2014})}\BibitemShut {NoStop}%
\end{thebibliography}%

\newpage

\pagebreak

\setcounter{equation}{0}
\setcounter{figure}{0}
\setcounter{table}{0}
\setcounter{page}{1}
\makeatletter
\renewcommand{\theequation}{S\arabic{equation}}
\renewcommand{\thefigure}{S\arabic{figure}}

\section{Supplementary Material for ``High-temperature Majorana zero modes''}

\subsection{Approximate analytic solution}

By appealing to the Bessel function identities $(\partial_x\pm
m/x)J_m(x)=\pm J_{m\mp 1}(x)$ it is easy to see that the solution to
Eqs.\ \eqref{h13} in the gapless limit $\delta=0$ takes the form
\begin{equation}\label{s1}
 \chi_0(\rho)=A\begin{pmatrix}
   J_1(\rho) \\
   J_2(\rho)
    \end{pmatrix}.
 \end{equation}
 For $x\gg 1$ the Bessel functions behave as
\begin{equation}\label{s2}
 J_m(x)\simeq \sqrt{2\over \pi x}\cos\left(x-{m\pi\over 2}-{\pi\over 4}\right),
 \end{equation}
which indicates that $\chi_0(\rho)$ is not normalizable, just as one
would expect of a scattering state in the absence of a gap.

Now consider the effect of turning on small nonzero $\delta$.  We
seek the solution in the form $f(\rho) \chi_0(\rho)$ where $f$ is a
slowly-varying envelope function.  In the
long-distance limit $\rho\gg 1$ the dominant contribution to the pairing
operator in Eqs.\ \eqref{h13} will then come from $\partial_\rho^2$
acting on $\chi_0(\rho)$ while the other terms will be suppressed
by powers of $1/\rho$. Furthermore, given the asymptotic form Eq.\
\eqref{s2} we may approximate in this limit $\partial_\rho^2
J_m(\rho)\approx -J_m(\rho)+ {\rm o}(\rho^{-1})$.  For small $\delta$
and large $\rho$  Eq.\ \eqref{h13} can thus be approximated as
\begin{eqnarray}\label{s3}
\chi_\uparrow&=&\left[\left(\partial_\rho+{2\over
                 \rho}\right)+\delta\right]\chi_\downarrow, \nonumber \\
  \chi_\downarrow&=&-\left[\left(\partial_\rho-{1\over \rho}\right)+
                     \delta\right]\chi_\uparrow.
\end{eqnarray}
The solution is an exponentially decaying oscillatory function
$\chi_\infty(\rho)$ given in Eq.\ \eqref{h14}. This
wavefunction is normalizable for any $\delta > 0$ and represents a
legitimate zero-energy eigenstate bound to a vortex defect in a gapped system. Perhaps
surprisingly we find that $\chi_\infty(\rho)$ is actually very close to the exact
solution obtained by numerical integration of full Eqs.\ \eqref{h13}
at all distances.

\subsection{Antivortex zero mode}

\begin{figure}[h]
\includegraphics[width = 8.6cm]{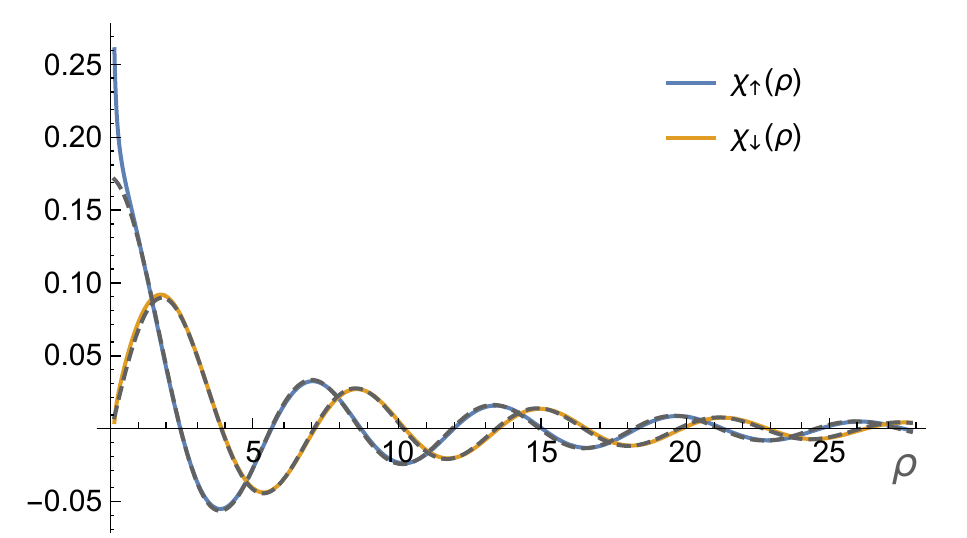}
\caption{Antivortex: Normalized zero-mode solutions for $\delta=0.06$ obtained by
  numerically integrating Eqs.\ \eqref{s4} (solid lines). Dashed lines
  represent the asymptotic forms given in Eq.\ \eqref{s5}.    
}\label{fig5}
\end{figure}

For an antivortex we consider Eq.\ \eqref{h12}  with $n=-1$ and
$l=0$. Working out the expression 
for the required pairing operators $\hat\Delta_{0,1}$ and
$\hat\Delta_{1,0}$  it becomes
\begin{eqnarray}\label{s4}
\chi_\uparrow&=&\left[\left(\partial_\rho+{1\over
                 \rho}\right)-\delta\left(\partial^2_\rho+{2\over\rho}\partial_\rho-{5\over
                 4\rho^2}\right)\right]\chi_\downarrow, \nonumber \\
  \chi_\downarrow&=&-\left[\partial_\rho-
                     \delta\left(\partial^2_\rho-{5\over 4\rho^2}\right)
  \right]\chi_\uparrow. 
\end{eqnarray}
The full solution, once again, must be obtained by numerical
integration; an example is shown in Fig.\ \ref{fig5}. The
long-distance asymptotic solution can be obtain by the same argument
as above and reads 
\begin{equation}\label{s5}
 \chi_\infty(\rho)=Ae^{-\rho\delta}\begin{pmatrix}
   J_0(\rho) \\
   J_1(\rho)
    \end{pmatrix}.
 \end{equation}
 As for the vortex we find that when the dimensionless gap $\delta$ is
 small this
 asymptotic solution turns out to approximate the exact numerical
 solution very well at all distances (see Fig.\ \ref{fig5} for an
 explicit comparison).

 It is interesting to note that the asymptotic solution  $\chi_\infty(\rho)$ in Eq.\
 \eqref{s5} has the same form as the exact zero-mode solution for a
 {\em vortex} in the Fu-Kane model at a non-zero chemical potential
 \cite{cheng2010tunneling}. In that model, however,
 $\delta=\Delta_0/\mu$, reflecting the fact that for an $s$-wave
order parameter the gap is simply $\Delta_0$.  Another difference is
that in the Fu-Kane model vortex and antivortex solutions are simply
related by time-reversal symmetry. In the present case time reversal
relates vortex in a $d+id'$ SC to an antivortex in a $d-id'$
SC. Therefore, there is no simple symmetry relation between vortex
and antivortex solutions in a $d+id'$ state, as can be seen e.g.\  by
comparing Eqs.\ \eqref{s5} and \eqref{h14} or the corresponding figures.

\subsection{Vortex core level spacing estimate}

According to the classic work of Caroli, de Gennes and Matricon
\cite{Caroli1964} a tower of bound states with energy spacing $\delta E\sim
\Delta^2/E_F$ is present in the vortex core of a conventional
$s$-wave BCS superconductor. We may expect such finite-energy CdGM
states to also exists in the present setup in addition to the Majorana zero
mode. Because the topological protection depends on the size of the
minigap (defined as the energy of the lowest excited state, $\delta E$
in our serup) it is important to estimate its magnitude.

As noted in Ref.\ \cite{Caroli1964} the vortex core can be regarded
as a disk of radius $R_v=c\xi$ in which the SC order has been locally
suppressed. Here $\xi=\hbar v_F/\pi\Delta$ is  the BCS coherence
length and $c$ a constant of order one. Correspondingly, the core
levels can be viewed as electron states of the underlying normal metal
confined inside this disk. In a conventional BCS superconductor the number 
of such states with energy inside the gap $\Delta $ can be
estimated as $N_{\rm CdGM}\simeq  \rho(E_F) \Delta\pi  R_v^2$,
where $\rho(E_F)=m/\pi\hbar^2$ is the density of states of the
underlying normal metal in two dimensions. Assuming equally spaced
core states the level spacing becomes
\begin{equation}\label{s10}
\delta E\simeq {\Delta\over N_{\rm CdGM}}={1\over \rho(E_F)\pi  R_v^2}={\pi^2\over 2c^2}{\Delta^2\over
  E_F}.
 \end{equation}
 This result agrees with the detailed calculation of Ref.\
 \cite{Caroli1964} if one takes $c=\pi/\sqrt{2}\simeq 2.22$. 
 In conventional superconductors $E_F$ typically exceeds $\Delta$ by two or three orders of magnitude making $\delta E$ too small for individual CdGM states to be experimentally resolved. In NbSe$_2$, for example, $\Delta\simeq 1.1$ meV while $E_F\simeq 350$ meV.  This makes the expected level spacing $\delta E\simeq 0.003$ meV, too small to be resolved by scanning tunneling spectroscopy (or any other probe).

 We may similarly estimate the core level spacing in the proximitized
 TI surface. The key difference here is the linear Dirac dispersion
 $\epsilon_\bk=\hbar v|\bk|$ which yields the normal-state
 density of states $\rho(\omega)=\omega/2\pi\hbar^2v^2 $. Assuming the
 chemical potential of the surface electrons $\mu\gtrsim \Delta_{\rm TI}$, where
 $\Delta_{\rm TI}$ is the proximity-induced gap, we may estimate the
 number  of the core states  $N_{\rm CdGM}\simeq  \rho(\mu)
 \Delta_{\rm TI}\pi  R_v^2$. From this, the level spacing can be deduced as
\begin{equation}\label{s11}
\delta E\simeq {\Delta_{\rm TI}\over N_{\rm CdGM}}={1\over
  \rho(\mu)\pi  R_v^2}={4}{\Delta_{\rm topo}^2\over
  \mu}\left({v\over v_F}\right)^2,
 \end{equation}
 where we have taken $c=\pi/\sqrt{2}$ as before.
 We note that the proximity-induced gap $\Delta_{\rm TI}$ has canceled
 out in this expression, the energy spacing being determined by the
 density of states $\rho(\mu)$ and the coherence length $\xi$. Because
 the SC order in the TI is induced by the proximity effect we expect
 that the vortex core size will be controlled by the coherence length
 $\xi$ of the superconductor. In a pure $d$-wave SC the coherence
 length is momentum-dependent, reflecting the momentum-dependent
 structure of the $d$-wave gap function. In the $d+id'$ phase that we
 assume occurs in the twisted cuprate bilayer the core size will be
 controled by the twist-induced minimum gap that we denote $\Delta_{\rm topo}$ (the
 notation here reflects the fact that $\Delta_{\rm topo}$
 protects the topological edge modes). We thus take $\xi=\hbar
 v_F/\pi\Delta_{\rm topo}$  as reflected in the last expression of
 Eq.\ \eqref{s11}.   For sufficiently small surface chemical potential
 $\mu$ (but still large compared to $\Delta_{\rm TI}$)  Eq.\ \eqref{s11}
indicates that minigap $\delta E$ can be large. As a concrete example
we consider  $\mu=100$meV, $\Delta_{\rm topo}=10$meV and $v/v_F=1$,
which yields  $\delta E\sim 4$meV. While there is a large uncertainty
associated with this estimate it nevertheless suggests that a situation with very few CdGM states inside the induced gap $\Delta_{\rm TI}$ can be readily
 achieved,  in contrast to the conventional SC case discussed above. (Note that $\Delta_{\rm TI}$  is some fraction of
 $\Delta_{\rm topo} $, depending on the nature of the TI/SC
 interface.) The above conclusion is in line with the results of our
 lattice simulation shown in
 Fig.\ 3  which indicates a small number of  CdGM states inside the gap,
 with the majority clustered close to  $\Delta_{\rm TI}$.

\end{document}